\documentclass[preprintnumbers,prd,onecolumn,floatfix,superscriptaddress, nofootinbib]{revtex4-2}

\usepackage{setspace}
\usepackage{graphicx}
\usepackage{subfig}
\usepackage{epsfig}
\usepackage{bm}
\usepackage{amssymb}
\usepackage{float}
\usepackage{amsmath}
\usepackage{dcolumn}
\usepackage[colorlinks]{hyperref}
\usepackage[usenames,dvipsnames]{color}
\usepackage{enumitem}

\hypersetup{ breaklinks=true, pdfstartview={FitH}, colorlinks=true, linkcolor=blue, citecolor=red, filecolor=magenta, urlcolor=blue, anchorcolor=green, linktocpage=true }

\def\doi{http://doi.org}

\newcommand{\be}{\begin{equation}}
\newcommand{\ee}{\end{equation}}
\newcommand{\beano}{\begin{eqnarray*}}
\newcommand{\eeano}{\end{eqnarray*}}
\newcommand{\ba}{\begin{eqnarray}}
\newcommand{\ea}{\end{eqnarray}}

\begin{document}

\title{A constrained cosmological model in $ f(R,L_m) $ gravity}
\author{J. K. Singh}
\email{jksingh@nsut.ac.in}
\affiliation{Department of Mathematics, Netaji Subhas University of Technology, New Delhi-110078, India}
\author{Shaily}
\email{shaily.ma19@nsut.ac.in}  
\affiliation{Department of Mathematics, Netaji Subhas University of Technology, New Delhi-110078, India}
\author{Ratbay Myrzakulov}
\email{ rmyrzakulov@gmail.com}  
\affiliation{Ratbay Myrzakulov Eurasian International Centre for Theoretical Physics, Astana, Kazakhstan.}
\author{Harshna Balhara}
\email{harshnabalhara24@gmail.com}  
\affiliation{Department of Mathematics, Netaji Subhas University of Technology, New Delhi-110078, India}

\begin{abstract}
\begin{singlespace}

\qquad In this article, we study the expanding nature of universe in the contest of $f(R,L_m)$ gravity theory, here $ R $ represents the Ricci scalar and $ L_m $ is the matter Lagrangian density. With a specific form of $ f(R,L_m) $, we obtain the field equations for flat FLRW metric. We parametrize the deceleration parameter in terms of the Hubble parameter and from here we find four free parameters, which are constraints and estimated by using $H(z)$, $Pantheon$, and their joint data sets. Further, we investigate the evolution of the deceleration parameter which depicts a transition from the deceleration to acceleration phases of the universe. The evolution behaviour of energy density, pressure, and EoS parameters shows that the present model is an accelerated quintessence dark energy model. To compare our model with the $ \Lambda $CDM model we use some of the diagnostic techniques. Thus, we find that our model in $ f(R,L_m) $ gravity supports the recent standard observational studies and delineates the late-time cosmic acceleration.

\end{singlespace}
\end{abstract}
 
\maketitle


\section{Introduction} 
The main objective of current and future cosmological investigations is to better understand the nature of cosmic acceleration by putting the cold dark matter ($\Lambda$CDM) standard cosmological model and its variations to the test \cite{SST:1998fmf,SCP:1998vns,Planck:2015fie,Planck:2018vyg}. Despite the fact that the cosmological constant $ \Lambda $ matches the data from observations well, it has two main flaws: the coincidence problem and the cosmological constant problem \cite{Copeland:2006wr}. One may create gravity theories that are equal to $\Lambda$CDM at the background level but display unique and exciting signs on perturbation dynamics by using the cosmological constant as the underlying source of the accelerated behaviour. In light of this idea, we will investigate if there is a gravity theory that possesses these qualities and can prove to be better than the $\Lambda$CDM paradigm. \\

One of the most straightforward options is to include an auxiliary function of the Ricci scalar $R$ in the action, which results in the $ f(R) $ theory \cite{Buchdahl:1970ynr,Kleinert:2000rt,Kerner:1982yg}. $f(R)$ theories of gravity are distinct among higher-order gravity theories in that they appear to be the only ones that can prevent the well-known and disastrous Ostrogradski instability \cite{Sotiriou:2008ve,delaCruz-Dombriz:2006kob, Loo:2022qdj}. 
When one quantizes the fluctuations of the scalar field in the background metric, one obtains the heavy scalar particles as well as the graviton in the case of $f(R)$ gravity. Given the size of the scalar particles in $f(R)$ gravity, the pressure may be negligible and the strength of the interaction between these scalar particles and the ordinary matter should be on par with that of gravitational forces. Such a scalar particle may therefore be a prime candidate for dark matter. Furthermore, because $f(R)$ theories do not contain ghosts, they can be selected so that the additional degrees of freedom compared to those of GR do not inescapably lead to serious viability issues. Additionally, it has been seen that some models display chameleon behaviour, enabling the theory to have cosmic effects that explain for the universe's current acceleration \cite{Carroll:2003wy,Capozziello:2006dj,Amendola:2006kh,Nojiri:2006gh}. Severe weak field restrictions in the Solar System regime appear to rule out the majority of the models put forth so far \cite{Chiba:2003ir,Erickcek:2006vf,Chiba:2006jp,Nojiri:2007uq,Olmo:2006eh}, while workable models do exist\cite{Hu:2007nk,Sawicki:2007tf,Amendola:2007nt}. Along with the solar system and equivalence principle restrictions on $f(R)$ gravity, observational signals of $f(R)$ dark energy models have been provided in 
 \cite{Tsujikawa:2007xu,Liu:2017xef,Capozziello:2007eu,Starobinsky:2007hu}. It has been discussed in \cite{Nojiri:2007as,Nojiri:2007cq,Cognola:2007zu} that there are more $f(R)$ models that unify early inflation and dark energy and pass local tests. To learn more about the many cosmological consequences of $f(R)$ gravity models, see references \cite{Santos:2007bs,Capozziello:2008qc,Nunes:2016drj}.\\

The explicit coupling of an arbitrary function of the Ricci scalar R with the matter Lagrangian density $L_m$ was included in the theory as a generalization of the $f(R)$ gravity theories \cite{Bertolami:2007gv}. This model was expanded to include the scenario of arbitrary couplings in both matter and geometry \cite{Harko:2008qz}. The non-minimal matter-geometry coupling's effects on cosmology and astrophysics have been thoroughly researched in \cite{Harko:2010vs,Harko:2010zi,Nesseris:2008mq,Faraoni:2007sn,Faraoni:2009rk}. Of all the gravitational theories built in Riemann space, the $ f(R,L_m) $ gravity theory can be viewed as their maximum extension \cite{Harko:2010mv}. In this gravity theory, test particle motion is non-geodesic and an additional force orthogonal to the four velocity vectors appears. Gonclaves and Moraes considered the $f(R,L_m)$ gravity when analysing cosmology from non-minimal matter geometry coupling \cite{Goncalves:2021vim}. Wu et al. studied Constraints of energy conditions and DK instability criterion on $f(R,L_m)$ gravity models \cite{Wu:2014yya}. Wang and Liao have recently investigated the energy conditions in $ f(R,L_m) $ gravity \cite{Wang:2012rw}.\\

\qquad The structure of the article is as follows: In Section II, we derive Einstein Fields equations for $ f(R,L_m) $ theory of gravity and take the parametrized deceleration parameter in terms of the Hubble parameter. In the next section, we constrain the best fit values of free parameters by using Hubble data set $ H(z) $ (77 points), Pantheon data set (1048 points) and joint data set $(H(z)+Pantheon)$. In Section IV, we investigate the evolution profile of various cosmological parameters and diagnostics techniques. Finally, in the last section, we summarize the outcomes of our work.     

\section{Einstein Field Equations}
The gravitational interactions in $f(R,L_m)$ modified theory of gravity can be written as \cite{Harko:2010mv}
\begin{equation}\label{1}
    A=\int f(R,L_m)\sqrt{-g}d^4x
\end{equation}
where  $f(R,L_m)$ is a random function of the Ricci scalar $R$ and the matter Lagrangian density term $L_m$. By taking the variation in the $L_m$ term we can obtain the energy momentum tensor of the matter and the outcome is 
\begin{equation}\label{2}
    T_{ij}=-\frac{2}{\sqrt{-g}} \frac{\delta(\sqrt{-g} L_m)}{\delta g^{ij}}=g_{ij}L_m - 2\frac{\partial L_m}{\partial g^{ij}},
\end{equation}
By taking the variation in action, Eq. (\ref{1}) leads the expression as

\begin{equation}\label{3}
    \delta A= \int \Bigg[f^R(R,L_m)\delta R+f^{L_m}(R,L_m) \frac{\delta L_m}{\delta g^{ij}} \delta g^{ij}-\frac{1}{2} g_{ij} f(R,L_m) \delta g^{ij} \Bigg] \sqrt{-g} d^4x,
\end{equation}
where $f^R=\frac{\partial f}{\partial R}$ and $f^{L_m}=\frac{\partial f}{\partial L_m}$. Variation in Ricci scalar can be calculated from the following manner
\begin{equation}\label{4}
    \delta R= \delta(g^{ij}R_{ij})=R_{ij}\delta g^{ij}+g^{ij}(\nabla_{k} \delta \Gamma^k_{ij}-\nabla_j \delta \Gamma^k_{ik} ),
\end{equation}
where $\nabla_k$ is the covariant w.r.t $g_{ij}$ and using the variated value of Christoffel symbol \textit{i.e.} $\delta \Gamma^k_{ij}=\frac{1}{2}g^{k\zeta}(\nabla_i \delta g_{j\zeta}+\nabla_j \delta g_{\zeta i}-\nabla_\zeta \delta g_{ij})$, Eq. (\ref{4}) leads to 
\begin{equation}\label{5}
    \delta R= R_{ij}\delta g^{ij}+ g_{ij}\nabla_i \nabla^i \delta g^{ij}-\nabla_i \nabla_j \delta g^{ij}.
\end{equation}
Therefore, from Eq. (\ref{3}) we obtain the Einstein field equations for $f (R, L_m)$ gravity as
\begin{equation}\label{6}
    f^R(R,L_m) R_{ij}+(g_{ij}\nabla_i \nabla^i - \nabla_i \nabla_j) f^R(R,L_m) -\frac{1}{2}[f(R,L_m)-f^{L_m}(R,L_m)L_m]g_{ij}=\frac{1}{2} f^{L_m}(R,L_m) T_{ij}.  
\end{equation}
Here, the energy momentum tensor for perfect fluid filled universe is written as $T_{ij}=(\rho+p)u_i u_j+pg_{ij}$, where p is used for isotropic pressure, $ \rho $ is matter energy density and the component of four velocities \textit{i.e.} $u_i=(1,0,0,0)$. We use the flat FLRW metric line element to study the present model and it can be expressed as

\begin{equation}\label{7}
    ds^2 = -dt^2 + a^2(t)(dx^2+dy^2+dz^2),
\end{equation}

we obtain the EFEs from Eq. (\ref{6}) as follows

\begin{equation}\label{8}
    3H^2 f^R+\frac{1}{2}(f-f^R R-f^{L_m} L_m)+3H\dot{f}^R=\frac{1}{2}f^{L_m}\rho,
\end{equation}
and
\begin{equation}\label{9}
    \dot{H} f^R+3H^2f^R-\ddot{f}^R-3H\dot{f}^R+\frac{1}{2}(f^{L_m} L_m-f)=\frac{1}{2}f^{L_m}p.
\end{equation}

In $ f(R,L_m) $ theory of gravity many models have been discussed and the motivation behind the investigation of our model is the related work of Harko \textit{et al.} \cite{Harko:2014gwa}. Taking this work into account, we consider  the following functional form as:  

\begin{equation}\label{10}
     f(R,L_m)=\frac{R}{2}+L_m^\lambda+\zeta,
\end{equation}
where $\lambda$ and $\zeta$ are arbitrary constants. 

The Einstein field Eqs. (\ref{8}) and (\ref{9}) in $ f(R,L_m) $ gravity takes the form as:

\begin{equation}\label{11}
    3H^2=\rho^\lambda(2\lambda-1)-\zeta
\end{equation}
\begin{equation}\label{12}
    4\dot{H}+9H^2=n\rho^{n-1}p+\beta-\rho^n(n-1)
\end{equation}

As the above two field equations have three unknowns namely $ H $, $ \rho $, and $ p $, therefore we need an additional constraint to get the solution. In this regard we have a crucial dimensionless element in comprehending the expansion scenario of the universe, that is deceleration parameter $ q $. The negative value of $ q $ indicates the accelerating universe and the positive value of $ q $ shows the deceleration phase of the universe. If $ q=0 $ then it shows a constant rate of expansion of the universe. Hence, with this motivation, we express the deceleration parameter in terms of the Hubble parameter $ H $ as \cite{Tiwari:2020ice}

\begin{equation}\label{13}
    q=\alpha-\frac{\beta}{H}
\end{equation}

where $\alpha$ and $\beta$ are model parameters with $\beta>0$. The scale factor $a$ is solved by using (\ref{13}) and the relation $ q=-\frac{a\ddot{a}}{\dot{a}^2 } $ as

\begin{equation}\label{14}
    a(t)=n(e^{\beta t}-1)^\frac{1}{1+\alpha}
\end{equation}

where $ n $ is the free parameter. Also Eq. (\ref{14}) leads to the following solution for the Hubble parameter

\begin{equation}\label{15}
    H=\frac{\beta e^{\beta t}}{(1+\alpha)(e^{\beta t}-1)}
\end{equation}

Since the scale factor $ a $ in terms of redshift $z$ is expressed as $\frac{a}{a_0}=\frac{1}{1+z}$ where $a_0=1$ is the present value. Thus, using this relation Hubble parameter in terms of redshift can be written as

\begin{equation}\label{16}
      H=\frac{\beta  \left((n (z+1))^{\alpha +1}+1\right)}{\alpha +1}.
\end{equation}

And so Eqs. (\ref{11}), (\ref{12}) and (\ref{16}) leads the values of $\rho$ and $p$ as

\begin{equation}\label{17}
    \rho=\left(\frac{(\alpha +1)^2 \zeta +3 \left(\beta +\beta  (n (z+1))^{\alpha +1}\right)^2}{(\alpha +1)^2 (2 \lambda -1)}\right)^{1/\lambda },
\end{equation}
and

\section{Statistical observation for model parameters} \label{sec:floats} 

\qquad In this section, we present the observational data and modality to calculate the free model parameters. To find the best fit value of model parameters, we use Hubble data set $ H(z) $ (77 points), Pantheon data set (1048 points) and joint data set $(H(z)+Pantheon)$. For this methodology, we can obtain the value of Hubble parameter from Eq. (\ref{16}) in terms of present age $(t_0)$ and $ H_0 $ as 

\begin{equation}\label{19}
    H=H_0(1-e^{-\beta t_0})[{n(1+z)}^{1+\alpha}+1],
\end{equation}
Where $ H_0=67.4 km/s/Mpc $ \cite{Planck:2018vyg} and we constraint remaining model parameters $ n $, $ \alpha $, $ \beta $ and $ t_0 $ using Markov Chain Monte Carlo (MCMC) method with emcee library in python for the mentioned data sets.   

\subsection{Hubble Data set}
\qquad In the present model, we have taken 77 points data set from $ H(z) $ assessment. $ H(z) $ data is obtained  from cosmic chronometers and very useful to examine the dark section of the universe \cite{Chimento:2007da}. By minimizing the value of chi-square, we can get the best fit results for the model parameters. Therefore, $ \chi^2 $ is written as

\begin{equation}\label{20}
\chi _{Hub}^{2}(n,\alpha,\beta,t_0)=\sum\limits_{i=1}^{77} \frac{[H(n,\alpha,\beta,t_0,z_{i})-H_{obs}(z_{i})]^2}{\sigma _{z_i}^2},
\end{equation}
where $ H(n,\alpha,\beta,t_0,z_{i}) $ and $ H_{obs}$ stands for the theoretical and observed values of Hubble parameter respectively and $ \sigma_{(z_{i}} $ indicates the standard deviation for each $ H(z_i) $. \\

\begin{figure}\centering
	\subfloat[]{\includegraphics[scale=0.54]{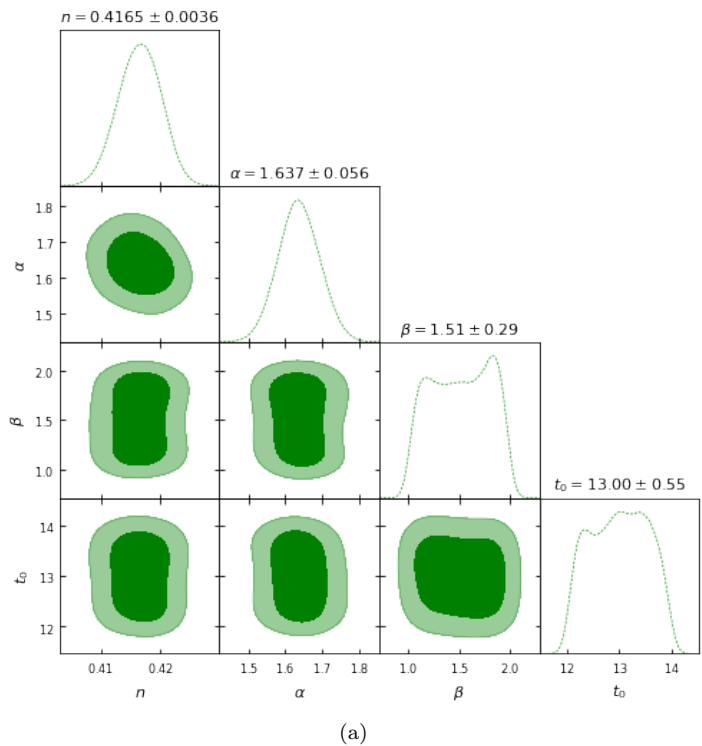}} 	
\caption{\scriptsize Likelihood contours for $ H(z) $ data set.}
\end{figure}

\begin{table}[H]
\caption{Constrained values of model parameters.}
\begin{center}
\label{tabparm}
\begin{tabular}{l c c c r} 
\hline\hline
\\ 
{Dataset} &      ~~~~~ $ n $  &  ~~~~~ $ \alpha $  & ~~~~~  $ \beta $ & ~~~~~ $ t_0 $  

\\
\\
\hline      
\\
{$ H(z) $ }     &  ~~~~~ $ 0.4165^{+0.0036}_{-0.0035} $   &  ~~~~~ $ 1.637^{+0.056}_{-0.056} $  &  ~~~~~ $ 1.51^{+0.29}_{-0.29} $   &  ~~~~~ $ 13.00^{+0.55}_{-0.55} $ 
\\
\\
{$ Pantheon $ }     &  ~~~~~ $ 2.1802^{+0.0013}_{-0.0013 } $   &  ~~~~~ $ 2.57^{+0.85}_{-0.85} $  &  ~~~~~ $ 3.9994^{+0.0006}_{-0.0002} $   &  ~~~~~ $ 13.12^{+0.82}_{-0.54} $ 
\\
\\
{ H(z)+Pantheon }  &  ~~~~~$ 0.4999^{+0.0001}_{-0.0001} $  & ~~~~~ $ 1.3019^{+0.0001}_{-0.0001} $  &  ~~~~~ $ 2.7284^{+0.0001}_{-0.0001} $   &  ~~~~~ $ 13.1006^{+0.0001}_{-0.0001} $
\\
\\ 
\hline\hline  
\end{tabular}    
\end{center}
\end{table}

\subsection{Pantheon Data}
\qquad In furtherance of study observational data and to constrain the free parameters namely $ n $, $ \alpha $, $ \beta $ and $ t_0 $, we use the currently accessible $ SNeIa $ sample, which is known as Pantheon sample \cite{Pan-STARRS1:2017jku}. In total Pantheon data contains 1048 points, which is the outcome of compilation of various $ SNeIa $ surveys. The CfA1-CfA4 surveys, the Carnegie Supernova Project (CSP), the Sloan Digital Sky Survey (SDSS), the Supernovae Legacy Survey (SNLS), the Pan-STARRS1 (PS1) and many Hubble Space Telescope (HST) samples incorporate in it \cite{Riess:1998dv, Jha:2005jg, Hicken:2009df, Contreras:2009nt, SDSS:2014irn}. In all these surveys, the range of redshift lies between 0.01 to 2.26. Supernovae data mainly use to investigate the expansion rate of the universe, thus for this probe we introduce the apparent magnitude as

\begin{equation}\label{21}
    m(z)=M+5log_{10}\Bigg[\frac{d_L(z)}{1Mpc}\Bigg]+25,
\end{equation}
where $ M $ is the absolute magnitude and $ D_L $ is the luminosity distance. For flat universe $ D_L $ can be obtain by 

\begin{equation}\label{22}
    d_L(z)=(1+z) \int_0^z \frac{c}{H(z^*)}dz^* ,
\end{equation}

where $ c $ is the speed of light. Also, the apparent magnitude can be expressed in terms of Hubble free luminosity distance $ (D_L(z)=\frac{H_0 d_L(z)}{c}) $ as

\begin{equation}\label{23}
    m(z)=M+5log_{10}[D_L(z)]+5log_{10}\Bigg[\frac{c/H_0)}{Mpc}\Bigg]+25.
\end{equation}

Here a degeneracy between $ M $ and $ H_0 $ can be notice and since distance modulus $ \mu =m-M$, therefore $ \chi^2 $ for Pantheon data can be written in terms of $ \mu $ as

\begin{equation}\label{24}
\chi_{SN}^{2}(n,\alpha,\beta, t_0)=\sum\limits_{i=1}^{1048}\left[ \frac{\mu_{th}(n,\alpha,\beta,t_0,z_{i})-\mu_{obs}(z_{i})}{\sigma _{\mu(z_{i})}}\right] ^2,
\end{equation}
where $ \mu_{th} $ and $ \mu_{obs} $  used for theoretical and observed distance modulus respectively. Here, we perform Markov chain Monte Carlo method to constrain the best fit value of free parameters, which are written in Table-I. \\

\begin{figure}\centering
	\subfloat[]{\includegraphics[scale=0.54]{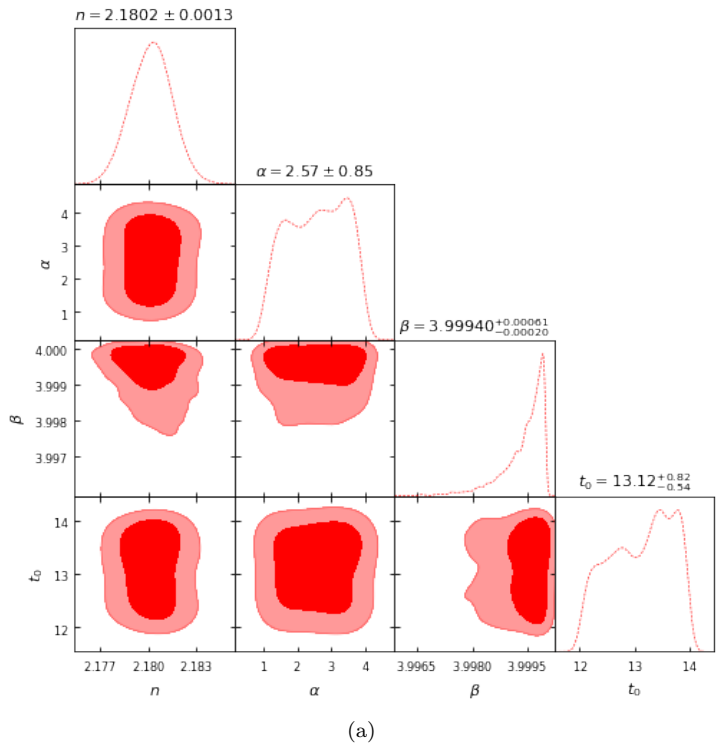}} 	
\caption{\scriptsize Likelihood contours for $ Pantheon $ data set.}
\end{figure}

\subsection{Joint Data sets ($ H(z)+Pantheon $)}

\qquad For joint analysis, the expression for $ \chi^2 $ function is given as
\begin{equation}\label{25}
\chi _{HS}^{2}(n,\alpha,\beta,t_0)=\chi _{Hub}^{2}(n,\alpha,\beta,t_0)+\chi _{SN}^{2}(n,\alpha,\beta,t_0).
\end{equation} 

\begin{figure}\centering
	\subfloat[]{\includegraphics[scale=0.54]{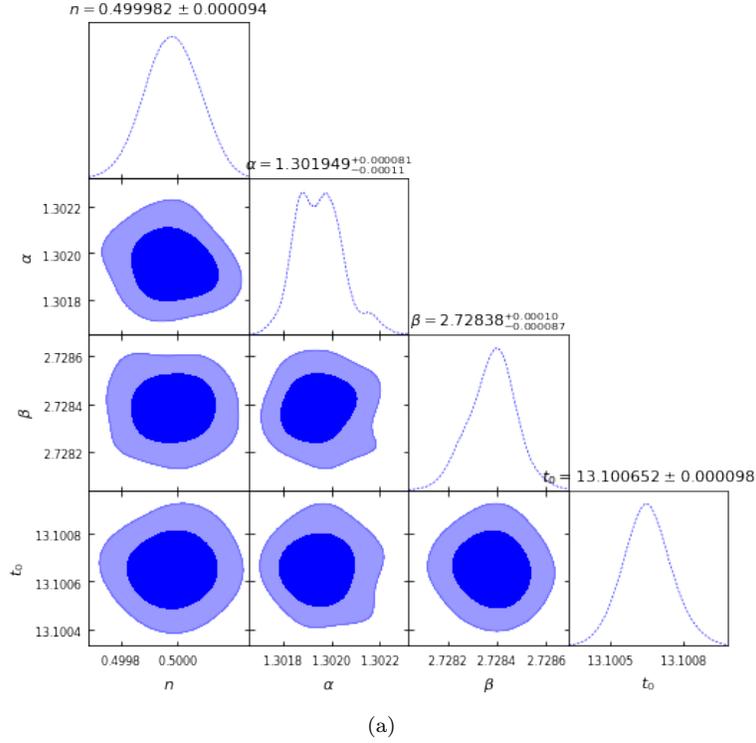}} 	
\caption{\scriptsize Likelihood contours for $ H(z)+Pantheon $ data set.}
\end{figure}

Joint data analysis is useful to get the stronger constraint value of the free parameters and the obtained values of $ n $, $ \alpha $, $ \beta $ and $ t_0 $, from all the observational data sets are placed in Table-I. \\

\section{Dynamical nature of the Universe}
 
\qquad In this section, we scrutinize various cosmological parameters with the constraint values of free parameters obtained from the Hubble data set, Pantheon data set and their combined data set. To understand the evolution of the universe in a better way, we study cosmic parameters one by one. From Eqs. (\ref{13}) and (\ref{16}), we can calculate the value of the deceleration parameter as
\begin{equation}\label{26}
    q=\alpha -\frac{\alpha +1}{(n (z+1))^{\alpha +1}+1}.
\end{equation}
From Fig. 4(a), we observe that the deceleration parameter changes its phase from positive to negative \textit{i.e.} universe transit from deceleration to acceleration. It is noticed that for the Hubble data set and joint data set, the universe was decelerating in the past and accelerating at present and in the future too. But the Pantheon sample shows deceleration in the past and present and afterwards, it accelerates in late times, which does not support recent evidence which suggests that the Universe's expansion is speeding up.

\begin{figure}\centering
	\subfloat[]{\label{a}\includegraphics[scale=0.46]{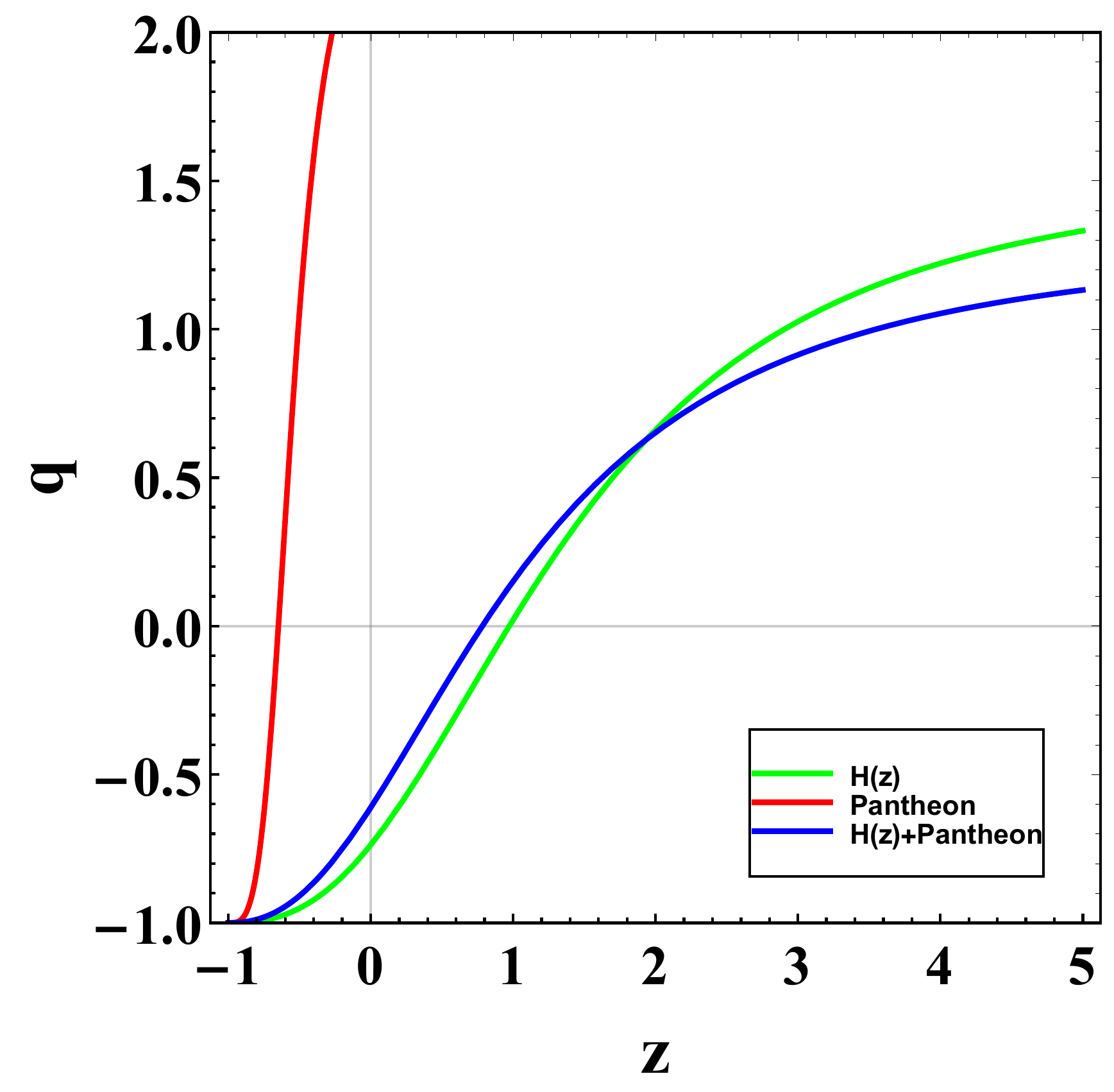}}\hfill
	\subfloat[]{\label{b}\includegraphics[scale=0.43]{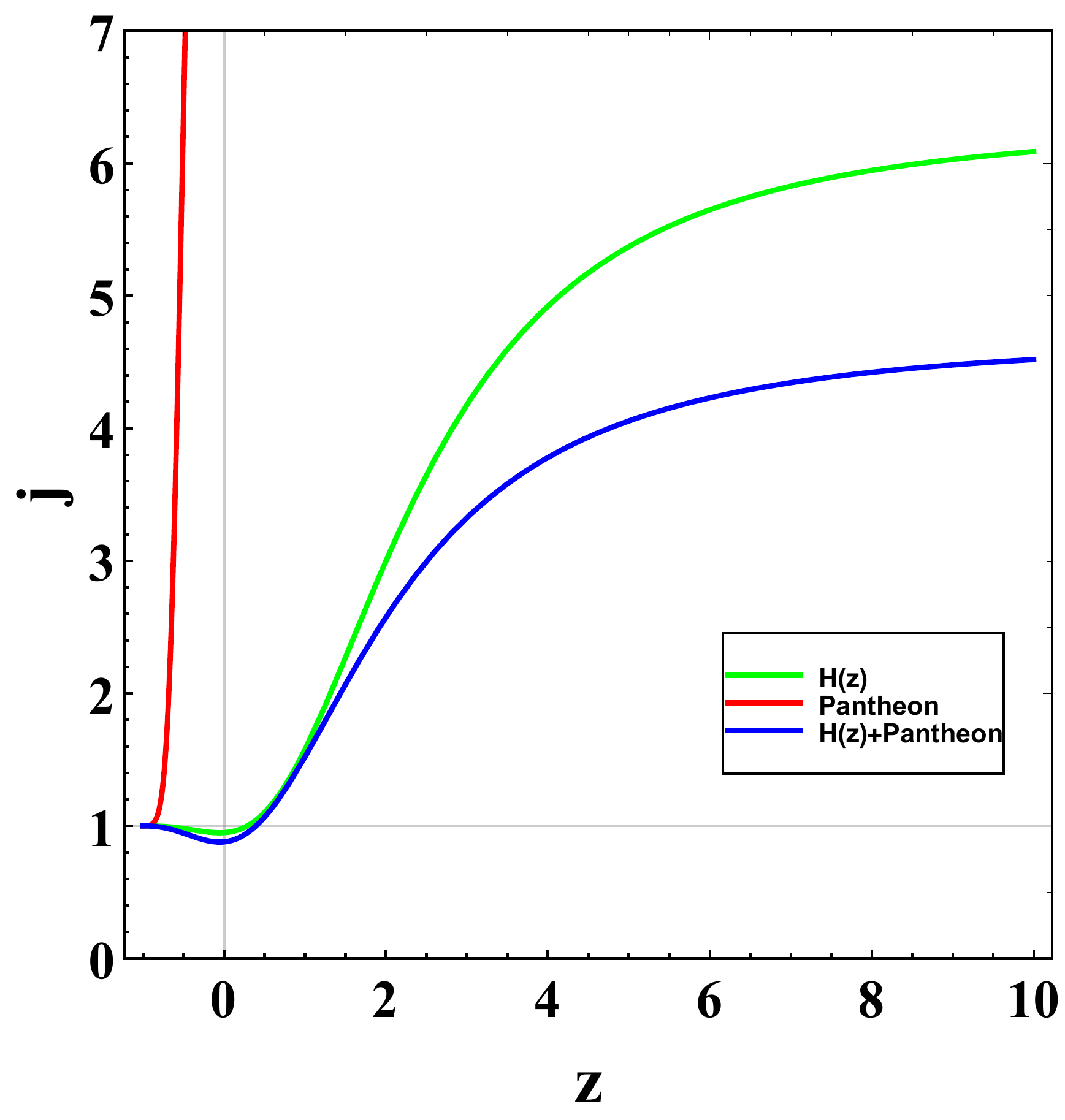}}\par
\caption{\scriptsize The graphs of $ q $ and $ j $ \textit{\ vs.} $ z $.}
\end{figure}

Now, to explore the nature of the universe in the present model, we study the parameters which contain the higher-order derivatives of $ a $. Here we discuss the jerk parameter ($ j $), which is also known as a jolt, pulse, bounce, impulse, surge \textit{etc.} \cite{Singh:2022eun}. The value of $ j $ can be calculated from the formula as 

\begin{equation}\label{27}
j=\frac{\dddot{a}}{a H^3},
\end{equation}
and using Eqs. (\ref{14}) and (\ref{16}), jerk parameter can be obtain in terms of redshift as
\begin{equation}
j=\frac{\alpha  \left(\alpha +(2 \alpha +1) (n (z+1))^{\alpha +1}-1\right) (n (z+1))^{\alpha +1}+1}{\left((n (z+1))^{\alpha +1}+1\right)^2}.
\end{equation}

Fig. 4(b) indicates that from early times to late times the value of $ j $ decreases for all observation samples and finally approaches to 1, which shows that this model is different from $ \Lambda CDM $ model at the early Universe and similar to $ \Lambda CDM $ model in late times.

To understand the physical behaviour of the universe we study various parameters like energy density $(\rho)$, isotropic pressure $(p)$, equation of state parameter $(\omega)$ etc. Using the expressions given in Eqs. (\ref{17}) and (\ref{18}), the plots of energy density and pressure are drawn in Fig. 5 and here we fixed the value of $ \lambda $ and $ \zeta $ as $ 0.55 $ and $ 20 $ respectively. Where the first figure shows that for all taken data sets, initially energy density is very high and afterwards it is receding from early times to late times. Fig. 5(b) depicts that for this model pressure is negative throughout the range of redshift for the mentioned data sets and according to the standard cosmology, negative pressure indicates the accelerated expansion of the universe. 

\begin{figure}\centering
	\subfloat[]{\label{a}\includegraphics[scale=0.45]{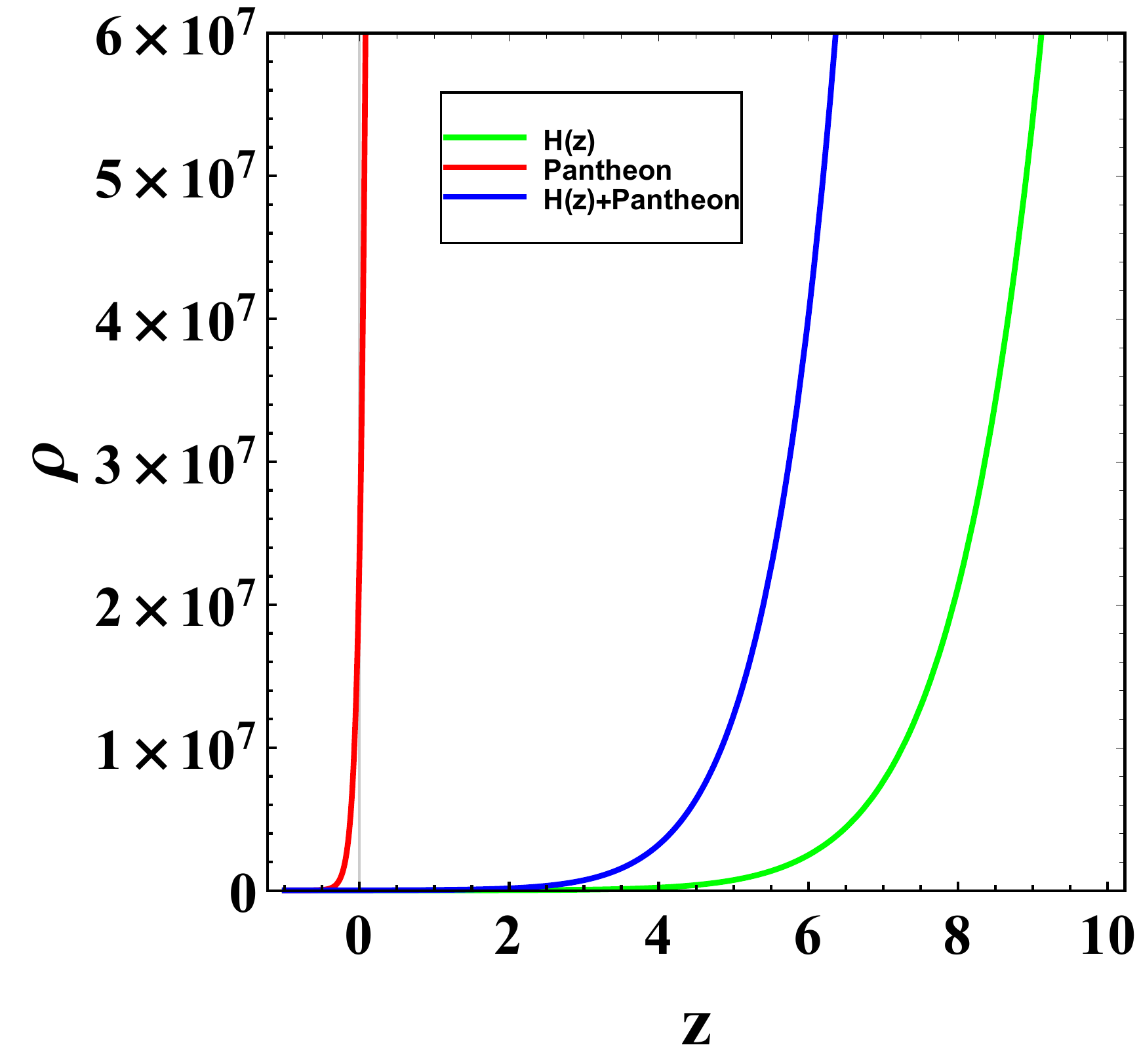}}	\hfill
	\subfloat[]{\label{b}\includegraphics[scale=0.47]{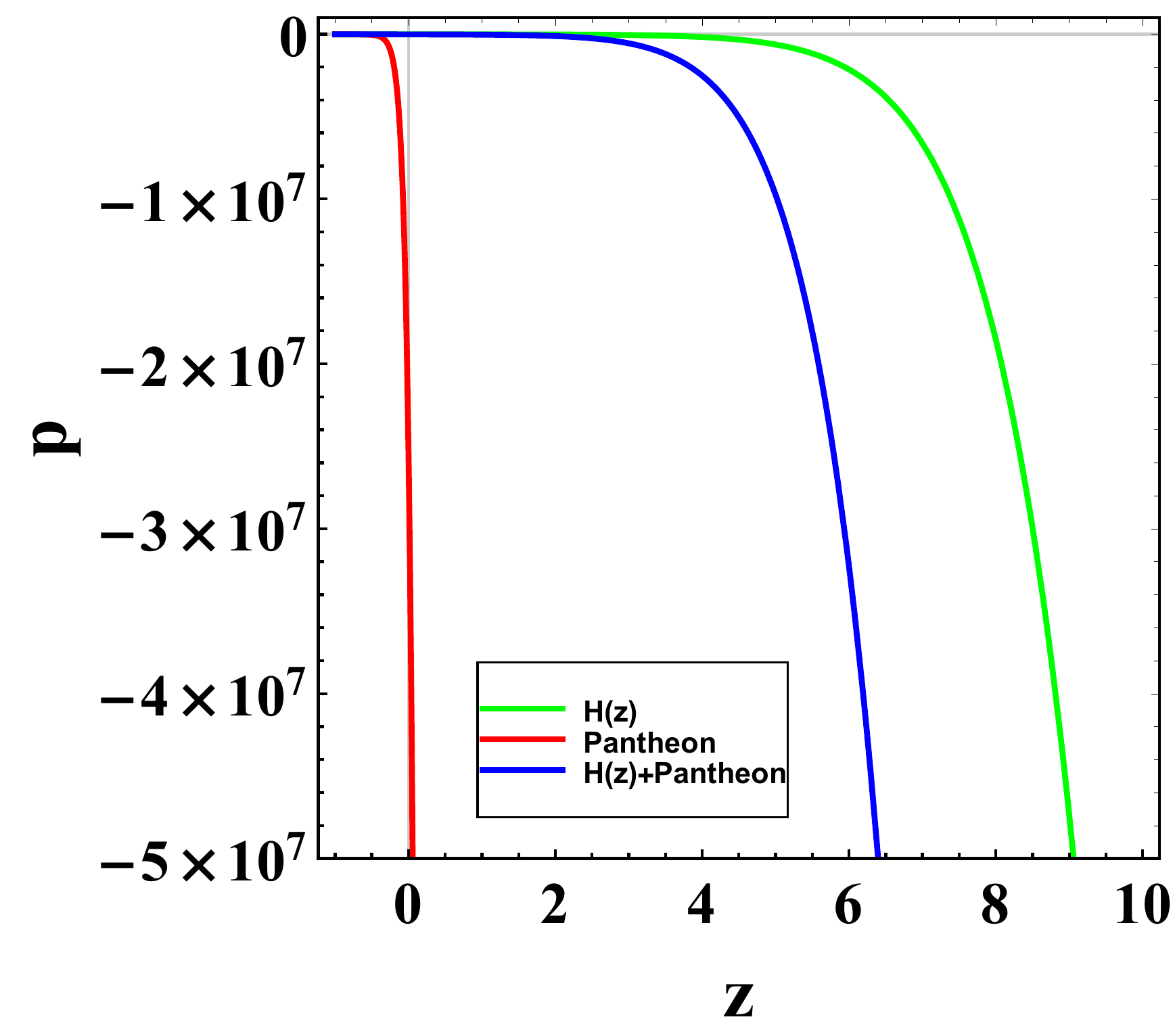}}\par
	\subfloat[]{\includegraphics[scale=0.45]{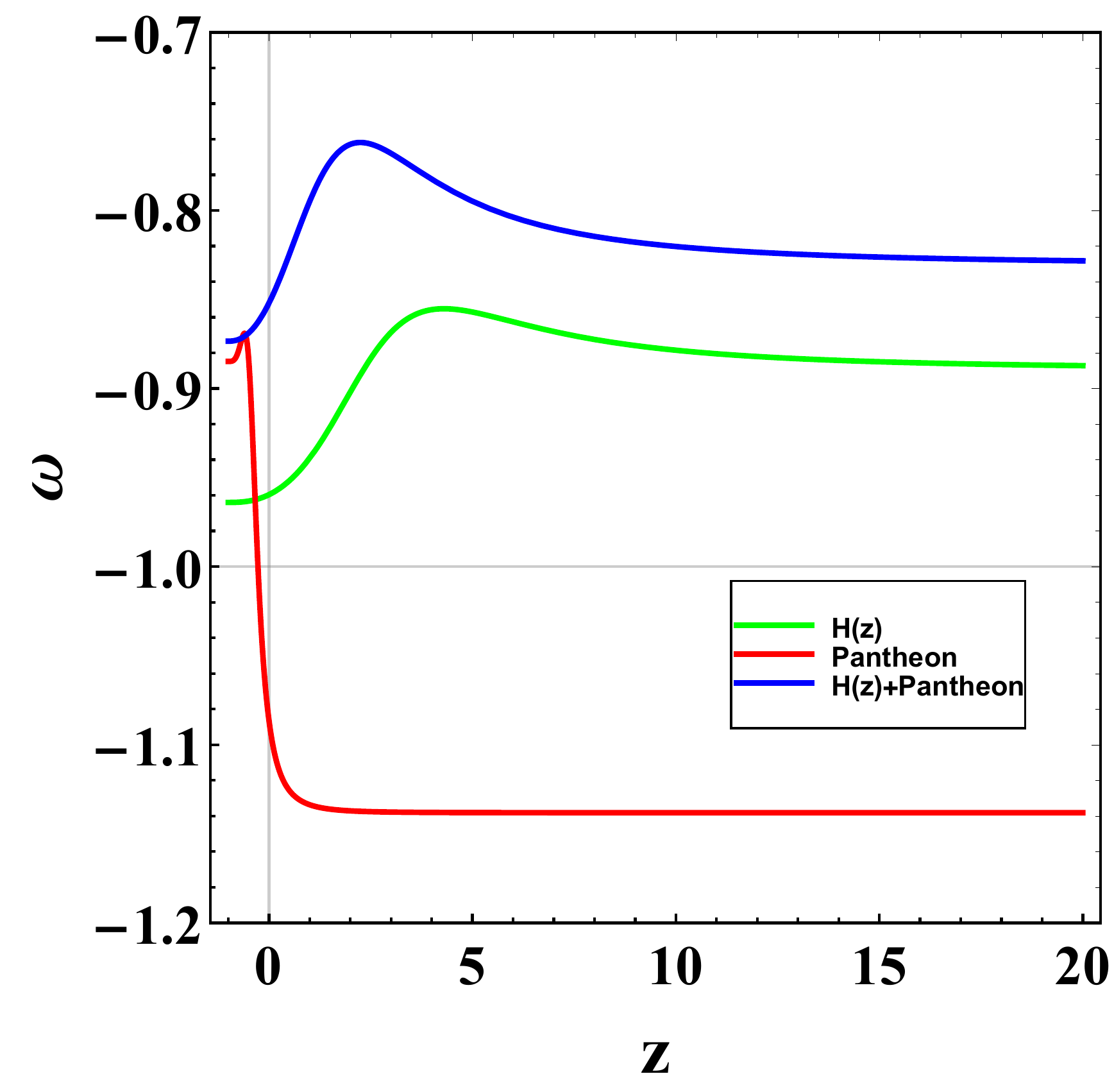}}
\caption{\scriptsize The graphs of $ \rho_{\phi} $, $ p_{\phi} $ and $ \omega $ \textit{\ vs.}  $ z $ for distinct observational data.}
\end{figure}

The value of the EoS parameter can be obtained from the equation $ \omega=\frac{p}{\rho} $ in terms of $ z $, where the value of $ p $ and $ \rho $ are taken from Eqs. (\ref{17}) and (\ref{18}). In Fig. 5(c), trajectories of $ \omega~ vs.~ z $ depict that for $ H(z) $ and joint data sets, the value of $ \omega $ lies in the quintessence region for the whole range of $ z $ and for $ Pantheon $ data set our model is in the phantom region at early times as well as at present but in future, it enters in quintessence region and remains there.

Now another tool to compare our model with $ \Lambda $CDM model and to understand the geometric behavior of the universe for the present model is statefinder diagnostic techniques \cite{Alam:2003sc}. In this technique, two geometrical diagnostic parameters $ (r,s) $ are introduced and formulated as:

\begin{equation}\label{28}
r=\frac{\dddot{a}}{aH^3},
\end{equation}

\begin{equation}\label{29}
s=\frac{r-1}{3(q-\frac{1}{2})}, ~~~where ~~ q\neq \frac{1}{2}  
\end{equation}

\begin{figure}\centering
	\subfloat[]{\label{a}\includegraphics[scale=0.28]{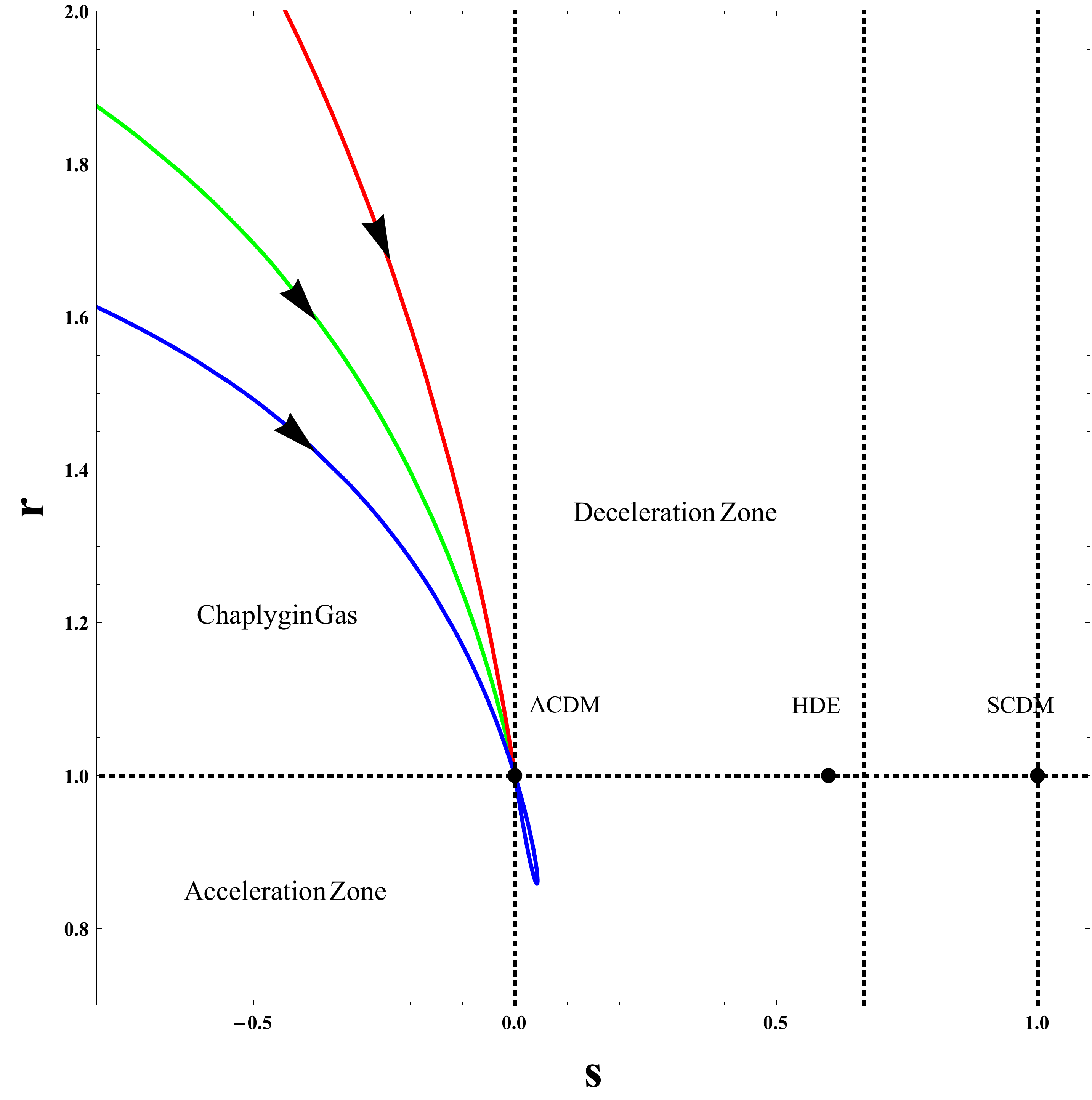}}\hfill
	\subfloat[]{\label{b}\includegraphics[scale=0.32]{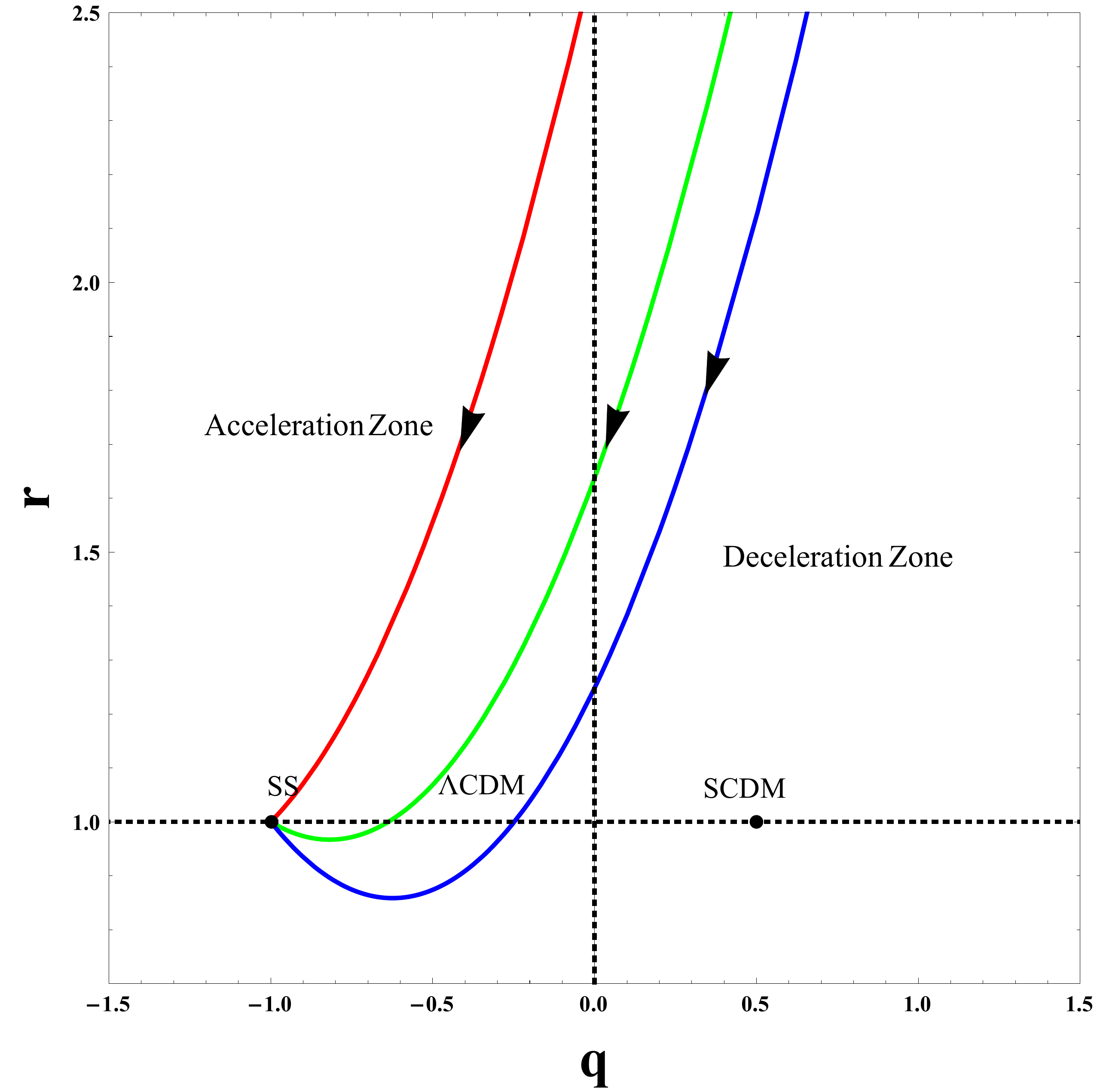}} 
	
\caption{\scriptsize The plots of $ s-r $ and $ q-r $.}
\end{figure}

Statefinder technique is useful to compare various dark energy model with $ \Lambda $CDM, $ SCDM $, $ HDE $ model. Now with the help of $ s-r $ trajectories it is quite easy to understand the nature of our model. In $ s-r $ plot, the point $ (0,1) $ represents the $ \Lambda $CDM model, $(2/3,1)$ denotes  holographic dark energy $(HDE)$ model and $(1,1)$ represents standard cold dark matter $ (SCDM) $ model. In our model, we plot the $ s-r $ curve by calculating the values of $ s $ and $ r $ from Eqs. (\ref{14}), (\ref{16}) and (\ref{26}). We notice that all the $ s-r $ trajectories are converging to $ \Lambda $CDM. As the arrows are indicating that trajectories for each observational data enters from Chaplygin Gas region and finally approach to $ \Lambda $CDM model. The present values of $(s,r)$ are $ (0.0079,0.9703) $, $(0.0424,0.8589) $, $ (2.3920,14.6384) $ for $ H(z) $, $ Pantheon $ and their joint data sets respectively, which shows that $ H(z) $, $ Pantheon $ and joint data trajectories are very close to $ \Lambda $CDM at present (see Fig. 6(a)).

In Fig. 6(b), we examine the $ q-r $ curves for each datasets, where the point $ (1,\frac{1}{2}) $ denotes the $ SCDM $ model, $ (-1,1) $ represents steady state $ (SS) $ model and the dotted horizontal line $ r=1 $ indicates $ \Lambda $CDM model. In our model, we observe that all the trajectories converge to $ SS $ model after crossing the transition line ($ \Lambda $CDM line). Thus, our model is similar to steady state model in late times as the evolution of the trajectories approaches to $ SS $ in future.

\qquad The Om diagnostic technique is generally used to observe the variations of various dark energy models from $ \Lambda $CDM model and  \cite{Sahni:2008xx, Zunckel:2008ti}. This tool can be discussed without using the EoS parameter and is defined in terms of $ z $ and $ H(z) $ as:
\begin{equation}
    Om=\frac{\Bigg(\frac{H(z)}{H_0}\Bigg)^2}{Z^3+3z^2+3z}.
\end{equation}
According to modern cosmology, positive curvature of the trajectory indicates the Phantom model, negative curvature of the trajectory shows the quintessence model and if there is no curvature then it indicates $ \Lambda $CDM model. In the present model, Fig. 7 shows that our model is a quintessence dark energy model for a large range of $z$.

\begin{figure}\centering
	\subfloat[]{\includegraphics[scale=0.5]{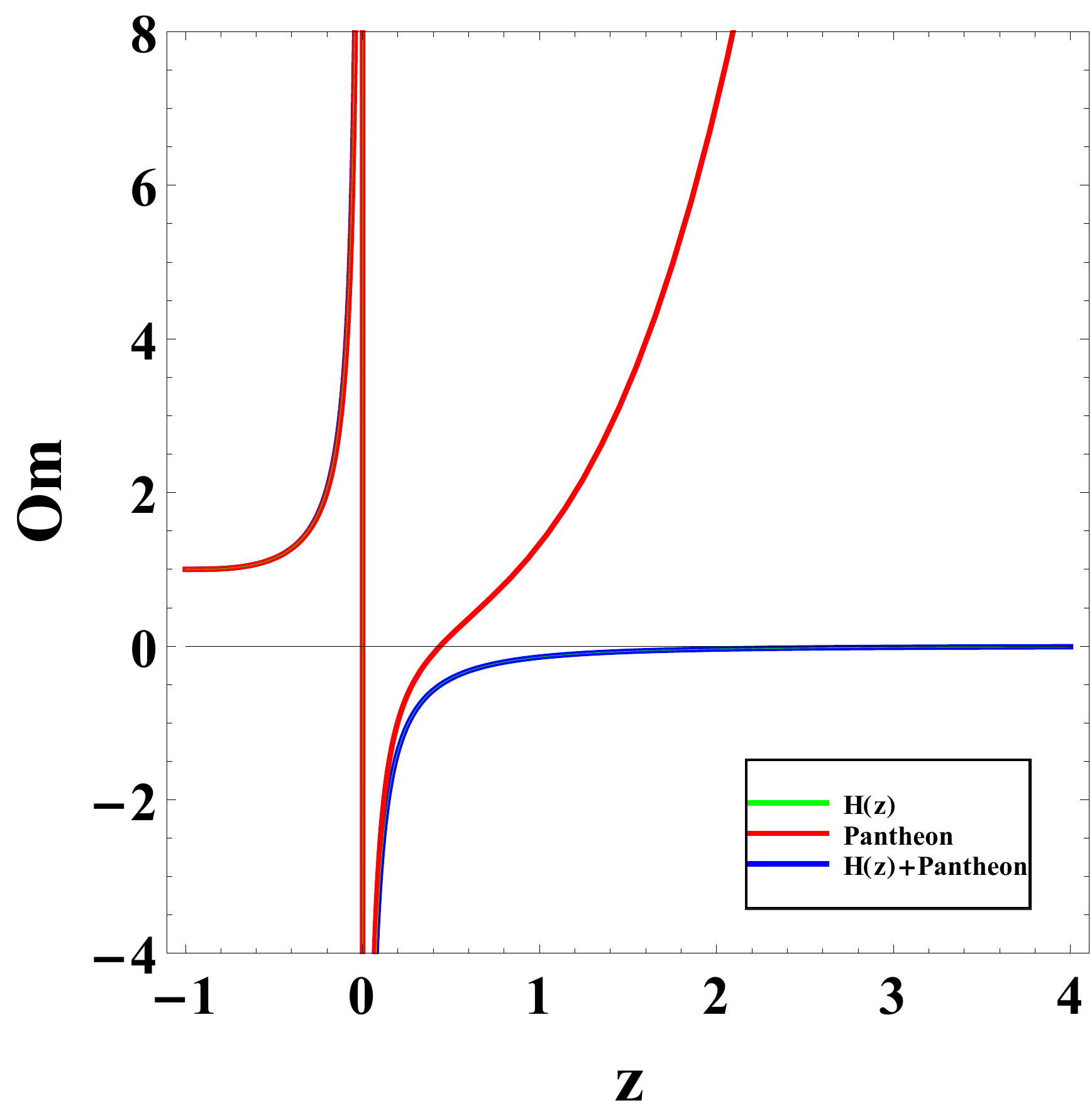}} 	
\caption{\scriptsize The plot of Om diagnostic \textit{\ vs.} $ z $.}
\end{figure}

\section{Conclusion}
In this article, we have studied the late-time behavior of the universe for a flat isotropic FLRW metric in $ f(R,L_m) $ theory of gravity. Now, to find a suitable solution set of the Einstein Field equations,  we parametrize the deceleration parameter in terms of Hubble parameter $ H $ and evaluate the scale factor $ a $ as a function of $ t $. We constrained the free parameters using Hubble $ H(z) $, $ Pantheon $, and their joint data sets. The best-fitted values of the free parameters are placed in Table-I. In addition, we have investigated the various other cosmological parameters using the observational values. At present, deceleration parameter shows the acceleration phase of the Universe for $ H(z)$ and joint data sets but for $ Pantheon $ data picture is little different (see Fig. 4(a)). At late times our model behaves like $ \Lambda $CDM and this is the outcome of jerk parameter as the value of j tending to $ 1 $ for all taken data sets. 

In this model, the energy density decreases monotonically as $ z $ decreases and at late times it diminishes to zero at late time. Isotropic pressure increased from high redshift to low redshift, and it is negative throughout the range of $ z $, which indicates the accelerating expansion of the universe. Also, the evolution profile of the EoS parameter $ \omega $ shows that the curves of $ \omega $ lie in the quintessence region for $ H(z)$ and joint data sets, and it  $ \omega $ changes its value from phantom to quintessence region for $ Pantheon $ data sets but all the three curves show the quintessence dark energy model at late time. Furthermore, we investigated statefinder diagnostics where the $ s-r $ plot also shows that at late times our model converges to $ \Lambda $CDM model and $ q-r $ trajectory represents that at the end our model converges to $ SS $ model. Finally, the evolution profile of the Om diagnostic indicates that our is a quintessence model.

\end{document}